\documentclass[
twocolumn,prl,]{revtex4-2}
\usepackage{amsmath}  
\usepackage{graphicx} 
\usepackage{physics}
\usepackage{amssymb}
\usepackage[final]{hyperref} 

\newcommand{\bx}{\mathbf{x}}
\newcommand{\bk}{\mathbf{k}}

\begin{document}

\title{\vspace*{0.5in} 
A First Analysis of Stochastic Composite Gravity
\vskip 0.1in}
\author{Joshua Erlich}\email[]{erlich@physics.wm.edu}
\affiliation{High Energy Theory Group, Department of Physics,
William \& Mary, Williamsburg, VA 23187-8795, USA}
\date{August 18, 2022}

\begin{abstract}
We present a first analysis of  a  nonperturbative approach to quantum gravity based on a representation of quantum field theory in terms of stochastic processes. The stochastic description accommodates a physical Lorentz-invariant ultraviolet regulator that provides a novel description of physics at ultra-short distances.
 In a stochastic composite gravity model we demonstrate the evolution of a generic initial field configuration towards an equilibrium in which the composite spacetime metric fluctuates about a flat spacetime.
We argue that fluctuations about the vacuum give rise to an emergent gravitational interaction consistent with Einstein gravity at long distances.  
We uncover a formal similarity between regularization by stochastic discreteness and point-splitting regularization in the corresponding quantum field theory. 
We comment on the  signature of the emergent spacetime,  possible consequences for the early universe, and the potential for observational and experimental tests of the stochastic origin of quantum field theory and gravitation.
\end{abstract}
\maketitle

\section{Introduction}

\vspace{-0.5\baselineskip}
The development of a quantum theory of gravity has been hindered by a number of conceptual and technical challenges stemming from diffeomorphism invariance, spacetimes with nontrivial causal structure, and the role of time as a parameter for the dynamics. A clue was provided by Sakharov's observation that diffeomorphism invariance should lead to the appearance of the Einstein-Hilbert action in the partition function of a generic regularized quantum field theory  in a background spacetime  \cite{Sakharov:1967pk,Visser:2002ew,Carlip:2012wa}. However, it has remained unclear what benefit might be drawn from this observation because the spacetime degrees of freedom would still require quantization.

Related to Sakharov's  observation is the paradigm of composite gravity, according to which there is no fundamental spacetime geometry, but instead the role of the  vielbein or metric is  played by a composite operator composed of the matter and gauge fields \cite{Amati:1981rf,Akama:1978pg,Guendelman:1996qy}. In composite gravity the gravitational interaction arises as an artifact of the ultraviolet regulator, so the regulator cannot be removed in a limit. Hence, identifying a physical regulator is critical to the success of the composite gravity program.
A number of analyses have confirmed the emergence of a gravitational interaction in toy models of composite gravity, albeit with traditional unphysical regulators such as dimensional regularization and Pauli-Villars fields which are generally believed to lead to inconsistencies as the regulator scale is approached \cite{Carone:2016tup,Carone:2017mdw,Chaurasia:2017ufl,Carone:2018ynf,Carone:2019xot,Batz:2020swk}.

In this paper we present a first analysis of a framework for quantum gravity based on a representation of quantum field theory in terms of stochastic processes. Discreteness of random fluctuations of the matter fields provides a physical ultraviolet regulator and a novel description of physics at ultrashort distances from which both quantum field theory and gravitation emerge as effective descriptions in an appropriate regime. The suggestion that quantum gravity might arise from a definition of quantum field theory in terms of stochastic processes is not novel~\cite{Markopoulou:2003ps,Calcagni:2017jtf,Moffat:1996fu,Erlich:2018qfc,Kuipers:2021jlh,Kuipers:2021aok}, but to the author's knowledge this is the first concrete realization of the paradigm in a composite gravity setting. 

\section{Composite Gravity}
There is a straightforward algorithm, at least in principle,  for constructing a composite gravity model whose effective description at long distances and weak fields is a specified quantum field theory coupled to gravity: We begin with the field theory coupled to an auxiliary spacetime background as in Sakharov's induced gravity scenario, and we replace the spacetime metric or vielbein (up to local Lorentz transformations) with the solution to the constraint $T_{\mu\nu}=0$, where $T_{\mu\nu}$ is the energy-momentum tensor in the curved-space background. Here we consider the scalar toy model investigated in Refs.~\cite{Akama:1978pg,Carone:2016tup}, which may be obtained from a theory of $N\gg1$  real scalar fields minimally coupled to an auxiliary metric $g_{\mu\nu}$ with action,
\begin{equation}
S_{{\rm aux}}=\int d^dx\sqrt{g}\left\{\frac12\sum_{a=1}^N\partial_\mu\phi^a\,\partial_\nu\phi^a\,g^{\mu\nu}-V(\phi^a)\right\}. \label{eq:Saux}
\end{equation}
The constraint of vanishing energy-momentum tensor determines the auxiliary metric as a function of the fields:
\begin{eqnarray}
T_{\mu\nu}=\sum_a\partial_\mu\phi^a\,\partial_\nu\phi^a -g_{\mu\nu}\Big(&&\frac12\sum_a\partial_\alpha\phi^a \,\partial_\beta\phi^a\,g^{\alpha\beta} \nonumber \\
&&-V(\phi^a)\Big)=0. \label{eq:Tmn}
\end{eqnarray}
There is an analytic solution to Eq.~(\ref{eq:Tmn}) for $g_{\mu\nu}$ in this toy model:
\begin{equation}
g_{\mu\nu}[\phi^a(x)]=\left(d/2-1\right)\frac{\sum_a\partial_\mu\phi^a\,\partial_\nu\phi^a}{V(\phi^a)} .\label{eq:compositeg}
\end{equation}
Eq.~(\ref{eq:compositeg}) defines the composite metric operator. Substituting the composite metric into the action Eq.~(\ref{eq:Saux}) gives the background-independent action that serves to define the theory:
\begin{equation}
S=\int d^dx\,\left(\frac{d/2-1}{V(\phi^a)}\right)^{\frac{d}{2}-1} \sqrt{\rm det\left(\sum_a\partial_\mu\phi^a\, \partial_\nu\phi^a\right)}. \label{eq:S}
\end{equation}
A fundamental spacetime metric does not appear in Eq.~(\ref{eq:S}), which we take as the starting point for the remainder of the analysis.

Expanding the composite operators in Eq.~(\ref{eq:S}) about a presumed vacuum expectation value, it is possible to compute scattering amplitudes by summing over the leading Feynman diagrams in a $1/N$ expansion. 
Such analyses have confirmed the emergence of a gravitational interaction in vacuum composite spacetimes  with and without cosmological constant \cite{Carone:2016tup,Carone:2017mdw,Chaurasia:2017ufl,Carone:2018ynf,Carone:2019xot,Batz:2020swk}. 
In all of these analyses, either dimensional regularization or Pauli-Villars fields were used as a proxy for the presumed physical regulator.The stochastic composite gravity framework completes the composite gravity scenario by  providing a physical ultraviolet regulator responsible for the emergent gravitational interaction.

\section{Stochastic Field Theory}
The framework introduces a discreteness into the stochastic description of quantum mechanics developed by Nelson \cite{Nelson:1983fp} and extended to quantum field theory by Guerra and Ruggiero \cite{Guerra:1973ck,Guerra:1981ie}, an idea proposed as a path towards quantum gravity in Refs.~\cite{Erlich:2018qfc}. The spacetime coordinates remain continuous, but the set of events at which the fields experience stochastic fluctuations is taken to be discrete. The ontology of the theory is straightforward, and includes only physical  fields whose dynamics include discrete random fluctuations. States are described by stochastic processes with generally nonlocal kinematics (which describes entanglement), but satisfying local dynamical equations of motion.

We first briefly review the stochastic description of free scalar quantum fields $\phi^a$.
In stochastic field theory, states correspond to stochastic processes described by a Langevin-type equation of the form,
\begin{equation}
d\phi^a(\mathbf{x}) = b_\bx[\{\phi^b(\mathbf{x})\},t]\,dt+\sqrt{2D} \,dW^a_{t}(\mathbf{x}),
\end{equation}
where the drift function $b_\bx$ is a possibly nonlocal function of the fields and $t$; the diffusion parameter $D$ is for now taken to be constant (but will later be effectively scale dependent); and $W_t$ is the continuum Wiener process with expectation values, \begin{eqnarray}
\mathbb{E}[dW^a_t(\mathbf{x})]&=&0, \nonumber \\
\mathbb{E}[dW^a_t({\mathbf{x}})\,dW^b_t({\mathbf{x}'})]&=&\delta^3({\mathbf{x}}-{\mathbf{x}'})\delta^{ab}\,dt. \label{eq:dW}\end{eqnarray}
We use the Ito convention, so that $dW_t$ at time $t$ represents a contribution to the change in the field during a step $dt$ to the {\em future} of $t$. We note that a primitive notion of causality is built into the language of stochastic processes in this way.

 It is convenient in the Lorentz-invariant theory to decompose the fields in a complete orthonormal basis of modes $u_\bk(\bx)$ with $\phi^a(x)=\sum_\bk \phi_\bk(t)\,u_\bk(\bx)$. The modes satisfy the stochastic differential equation,
\begin{equation}
d\phi^a_\bk = b^a_\bk[\{\phi^b\},t]\,dt+\sqrt{2D} \,dW^a_\bk, \label{eq:dphi}
\end{equation}
where $\mathbb{E}[dW_\bk^a]=0$ and at each time $t$ \begin{equation}
\mathbb{E}[dW^a_\bk\,dW^b_{\bk'}]=\delta_{\bk\bk'}\delta^{ab}\,dt. \label{eq:dWq}
\end{equation}
Following Nelson we also consider the backwards process:
\begin{equation}d\phi^a_\bk = b^a_{*\bk}[\{\phi^b\},t]\,dt+\sqrt{2D} \,dW^a_{*\bk},\end{equation}
with $dt<0$ representing evolution backwards in time, and $dW^a_{*\bk}$  distributed as in Eq.~(\ref{eq:dWq}) but representing changes to the {\em past} of $t$. Note that in general $b_{*\bk}^a\neq b_\bk^a$.

Due to the nondifferentiable stochastic evolution in time, dynamics is most easily expressed in terms of smooth expectation values.
The stochastic forward and backward derivatives at time $t$, denoted $D_+$ and $D_-$, respectively, are defined in terms of expectation values over filtrations with $\phi(\mathbf{x};t)$ fixed at time $t$: \begin{eqnarray}
D_+ f(\phi):= \lim_{\Delta t \rightarrow 0}\mathbb{E}\left[\frac{f(\phi(t+\Delta t))-f(\phi(t))}{\Delta t}\right]&\hspace{22pt}& \\
D_-f(\phi):= \lim_{\Delta t \rightarrow 0}\mathbb{E}\left[\frac{f(\phi(t))-f(\phi(t-\Delta t))}{\Delta t}\right]&& \label{eq:StochasticFreeEOM}
 \end{eqnarray}
 Unlike for classical functions, in general $D_+f\neq D_-f$.
The modes satisfy a generalization of the free-field equations of motion suitable for stochastic dynamics \cite{Guerra:1973ck,Guerra:1981ie}
\begin{equation}
\frac{1}{2}\left(D_+D_-\phi^a_\bk+D_-D_+\phi^a_\bk\right) +\omega_\bk^2\,\phi^a_\bk=0, \label{eq:flatEOM} \end{equation}
with $\omega_\bk^2=\left(\bk^2+m^2/\hbar^2\right)$.
The combination of $D_+$ and $D_-$ follows from time reversibility of the resulting dynamics \cite{Nelson:1983fp}.
Evaluating the stochastic derivatives by expanding the field about time $t$, one can write explicit expressions for the terms in Eq.~(\ref{eq:flatEOM}):
\begin{eqnarray}
D_+&=&\frac{\partial}{\partial t}+\sum_{a,\bk} \,\left\{b^a_\bk\frac{\partial}{\partial\phi^a_\bk}+D\,\frac{\partial^2}{\partial\phi_\bk^{a\,2}}\right\}, 
\nonumber \\
D_-&=&\frac{\partial}{\partial t}+\sum_{a,\bk} \,\left\{b^a_{*\bk}\frac{\partial}{\partial\phi^a_\bk}-D\,\frac{\partial^2}{\partial\phi_\bk^{a\,2} }\right\}.\end{eqnarray}

Unlike for classical functions, for  functions of stochastic variables the forward and backward derivatives are generally unequal. We note that \begin{equation}
D_+\phi^a_\bk=b^a_\bk, \ \ 
D_-\phi^a_\bk:=b^a_{*\bk}. \end{equation}
The convective velocity $v_\bk^a$ and stochastic velocity $u_\bk^a$ are defined as, \begin{equation}
v_\bk^a:=\frac{b_\bk^a+b_{*\bk}^a}{2}, \ \ \ 
u_\bk^a:=\frac{b_\bk^a-b_{*\bk}^a}{2}.\label{eq:vkuk} \end{equation}
By an H-theorem for stochastic processes \cite{H-theorem}, initial ensembles generically evolve towards a common probability distribution, the quantum equilibrium distribution satisfying the functional Schr\"odinger equation and the Born rule, with the diffusion parameter now identified with $D=\hbar/2$.

Conversely, given a solution to the functional Schr\"odinger equation $\psi[\phi(\bx),t]=\sqrt{\rho}e^{iS}$, the stochastic process whose probability density functional agrees with the Born rule $\rho=|\psi|^2$ for that state has convective and stochastic velocities given by
\cite{Nelson:1983fp},
\begin{equation}
v^a=2D\frac{\delta S}{\delta \phi^a}, \ \ \ 
u^a=D\frac{\delta\ln\rho}{\delta\phi^a}. \label{eq:vu} \end{equation}
With knowledge of the convective and stochastic velocities, the forward drift coefficient is then determined by 
$b^a=v^a+u^a$.
The stochastic process corresponding to the ground state of the free field is then found to be an Ornstein-Uhlenbeck type process \cite{Guerra:1973ck}, 
\begin{equation}
d\phi^a(\bx)=-\sqrt{-\nabla^2+m^2/\hbar^2}\phi^a(\bx)\,dt+\sqrt{2D}\,dW^a(\bx), \label{eq:dphi-vacx} \end{equation}
or in terms of the mode expansion,
\begin{equation}
d\phi^a_\bk=-\sqrt{\bk^2+m^2/\hbar^2}\phi^a_\bk\,dt+\sqrt{2D}\,dW_\bk^a. \label{eq:dphi-vac}\end{equation}
From Eqs.~(\ref{eq:vkuk}) and (\ref{eq:vu}), stationary states in equilibrium, including the vacuum, have $b_{*\bk}^a=-b_\bk^a$. From Eq.~(\ref{eq:dphi-vac}) we have $b_{\bk}^a=-\sqrt{\bk^2+m^2/\hbar^2}\phi_\bk^a$ in the vacuum state.
\section{Stochastic Composite Gravity}
Taking $V(\phi^a)=V_0+\sum_a m^2/2\,:\phi^a\phi^a:$ and $d=4$, where the normal-ordering indicates that the expectation value of the mass term is absorbed in the definition of $V_0$, in the stochastic description the action Eq.~(\ref{eq:S}) is replaced by,
\begin{equation}
S=\int d^4x\,\mathbb{E}\left[\frac{\sqrt{\det (V_0 G_{\mu\nu})}}{V(\phi^a)}\right], \label{eq:StochasticS}
\end{equation}
where, \begin{eqnarray}
&G_{00}&=\frac{1}{2V_0}\sum_a\left(D_+\phi^a\,D_+\phi^a +D_-\phi^a\,D_-\phi^a\right)
\nonumber  \\
&G_{i0}&=G_{0i}=\frac{1}{2V_0}\sum_a\left( D_+\phi^a+D_-\phi^a\right)\partial_i\phi^a \nonumber \\
&G_{ij}&=\frac{1}{V_0}\sum_a\partial_i\phi^a\,\partial_j\phi^a. \label{eq:Gmn}
\end{eqnarray}
With the approximation $V_0\gg m^2:\phi^2:$, at lowest order in $1/V_0$ we have $g_{\mu\nu}\approx G_{\mu\nu}$.
The equations of motion follow from stationarization of the action over stochastic processes, following the stochastic calculus of variations developed by Yasue \cite{Yasue}: \begin{equation}
D_+\frac{\delta S}{\delta (D_-\phi^a)}+D_-\frac{\delta S}{\delta (D_+\phi^a)}+\partial_i \frac{\delta S}{\delta (\partial_i\phi^a)}-\frac{\delta S}{\delta\phi^a}=0. \label{eq:SCG-EOM}\end{equation}
A version of Noether's theorem applied to stochastic mechanics \cite{Yasue} allows us to identify a conserved energy and momentum. Defining ${\cal L}$ through $S=\int d^4x{\cal L}$, the energy-momentum tensor can be identified as:
\begin{eqnarray}
T_0^{\,0}&=&\mathbb{E}\left[\sum_a \left(D_+\phi^a\frac{\partial{\cal L}}{\partial (D_+\phi^a)}+D_-\phi^a\frac{\partial{\cal L}}{\partial (D_-\phi^a)}\right)-{\cal L}\right] \nonumber \\
T_0^{\,i}&=&\mathbb{E}\left[\sum_a \left(D_+\phi^a\frac{\partial{\cal L}}{\partial (\partial_i\phi^a)}+D_-\phi^a\frac{\partial{\cal L}}{\partial (\partial_i \phi^a)}\right) \right]\nonumber \\
T_i^{\,0}&=&\mathbb{E}\left[\sum_a \left(\partial_i\phi^a\frac{\partial{\cal L}}{\partial (D_+\phi^a)}+\partial_i\phi^a\frac{\partial{\cal L}}{\partial (D_- \phi^a)}\right)\right] \nonumber \\
T_i^{\,j}&=&\mathbb{E}\left[\sum_a\left(\partial_i\phi^a \frac{\partial{\cal L}}{\partial (\partial_j\phi^a)}\right)-\delta_i^j\,{\cal L}\right].
\end{eqnarray}
It is straightforward to demonstrate that independent of the equations of motion, $T_\mu^{\,\nu}=0$  in the model under consideration, confirming a familiar consequence of diffeomorphism invariance. Solving the equations of motion Eq.~(\ref{eq:SCG-EOM}) is then analogous to solving the Wheeler-DeWitt equation of canonical quantum gravity \cite{DeWitt:1967yk}. 

By a coordinate reparametrization we choose to arrange stochastic events via a Poisson distribution per unit Cartesian volume, which then defines a local reference frame. The Poisson distribution defines the regularization length $\tau$. Poisson sprinklings (in infinite volume) respect Lorentz invariance  but not Euclidean invariance, because with a Euclidean metric the displacement of nearest neighbors determines a preferred direction \cite{Bombelli:1987aa,Christ:1982zq}. 
\begin{figure}[t] 
      \includegraphics[width=.5\columnwidth]{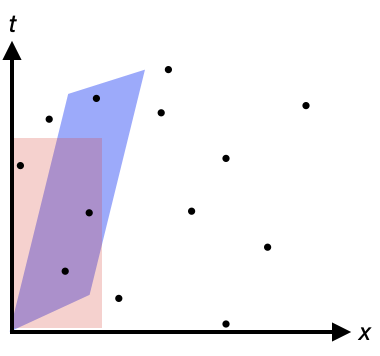}
\caption{The density of stochastic events in a Poisson sprinkling is Lorentz invariant.} \label{fig:Poisson}
\end{figure}

In terms of the microscopic description, Eq.~(\ref{eq:dphi})  does not give a precise description of the field evolution, but instead the Brownian contribution is to be replaced by a discrete set of  random $\delta$-function fluctuations at the Poisson sprinkling of events. A dynamical consequence of this is that the field in the equilibrium vacuum state has support  localized around the discrete set of stochastic events, resulting in an effective spacetime discreteness. 
Eq.~(\ref{eq:dphi-vacx}) then describes the evolution of the field after averaging over a small three-volume as it traces out a 4-volume that includes a large number of  stochastic events. 
 The density of stochastic events per unit four-volume is Lorentz invariant, as illustrated in Fig.~\ref{fig:Poisson}, so Eq.~(\ref{eq:dphi-vacx}) takes the same form in a Lorentz-transformed frame, with $\bx$ and $t$ transformed appropriately.  
 
In a mean field approximation in the continuous stochastic limit, we assume a vacuum described approximately by the Lorentz-invariant vacuum process Eq.~(\ref{eq:dphi-vac}), and we tune the constant $V_0$ in the potential so that $\mathbb{E}[G_{\mu\nu}]$ is properly normalized.

To this end, we calculate $\mathbb{E}\left[G_{\mu\nu}\right]$ for the process Eq.~(\ref{eq:dphi-vac}) in equilibrium. By Ito's lemma, we have,\begin{eqnarray}
d\mathbb{E}[(\phi_\bk^a)^2]&=&2\,\mathbb{E}[\phi_\bk^a\,d\phi_\bk^a]+2D\,dt \nonumber \\
&=&-2\omega_\bk \mathbb{E}[(\phi_\bk^a)^2]\,dt+2D\,dt,
\end{eqnarray}
which can be integrated to give, \begin{equation}
\mathbb{E}[(\phi_\bk^a)^2]=\frac{D}{\omega_\bk}+Ce^{-2\omega_\bk t}. \label{eq:Ephisq}
\end{equation}
The constant $C$ depends on the initial field configuration, but independent of $C$ the expectation value approaches a common equilibrium value. 
Similarly, we find that in the infinite-volume, continuous process limit, the composite metric approaches an equilibrium with\begin{equation}
\mathbb{E}\left[G_{\mu\nu}\right]=\frac{N}{V_0}\int \frac{d^3k}{(2\pi)^3}\frac{D \,k_\mu k_\nu}{\omega_\bk}. \label{eq:Gmn-vac}
\end{equation}
Taking $D=\hbar/2$ we recognize this formally as the one-loop vacuum expectation value of the composite operator $G_{\mu\nu}$ in the equivalent Lorentz-invariant quantum field theory. The UV divergence is regularized by discreteness of the stochastic process.  
As the discreteness scale is approached, Eq.~(\ref{eq:dphi})  no longer provides an accurate description of the kinematics as the diffusive effects diminish. 
This can be modeled by replacing the diffusion parameter $D$ with a scale-dependent $D(\omega_\bk)$. 
 If we take the typical discreteness scale in the Poisson sprinkling to be $\tau\ll \hbar/m$, then a natural choice for $D(\omega_\bk$) would be $
D(\omega_\bk)=e^{-\omega_\bk \tau}$, 
though we should average over timelike directions to capture the Lorentz invariance of the description.

\begin{figure}[ht] 
      \includegraphics[width=1\columnwidth]{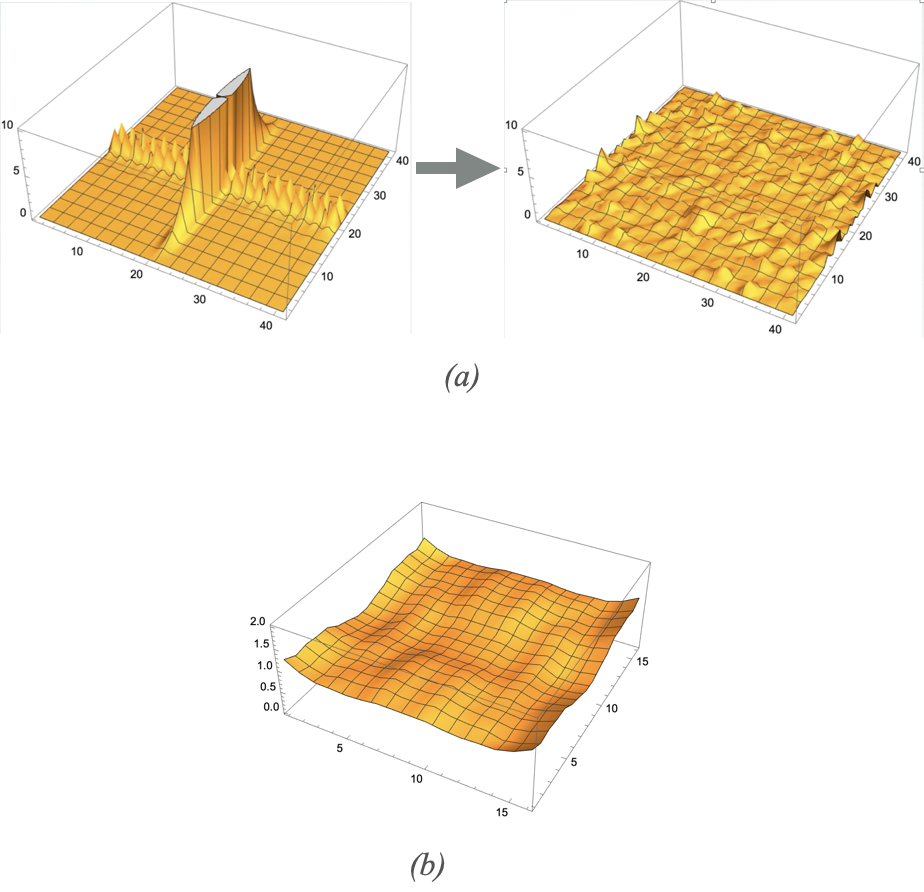}
\caption{a) Evolution of the composite operator $G_{00}$ from a far-from-equilibrium initial field configuration over $10^3$ time steps in 2+1 dimensions including the lowest $20\times20$ even-parity modes. b) $G_{00}$ in the final configuration from (a), but averaged around each point over $5\times5$ spatial volumes in units of the time step. $V_0$ can be tuned so that the uniform $G_{00}(\bx)$ in equilibrium is normalized to 1 self-consistently with the presumed Minkowski space vacuum.} \label{fig:G00}
\end{figure}

With this replacement of $D$ in Eq.~(\ref{eq:Gmn-vac}), the spacetime discreteness acts as a Euclidean-time point splitting regulator of the one-loop integral in the effective quantum field theory. We recall that  correlation functions of products of fields in stochastic field theory have a direct physical interpretation and are equivalent to Wick rotations of the Wightman functions in the corresponding quantum field theory, as was noted by Guerra and Ruggiero \cite{Guerra:1973ck,Guerra:1981ie}. Hence, we can interpret the above result as equivalent to the Lorentzian point splitting of the composite operator $G_{\mu\nu}$ in the quantum field theory.
Fig.~\ref{fig:G00}a illustrates a simulation of $G_{00}$  with a far-from-equilibrium initial field configuration and demonstrates the evolution towards a spatially uniform equilibrium configuration.

We see that locally there are large fluctuations of the composite metric even after equilibrium is reached, but $G_{00}$ appears uniform when averaged over regions of several volumes associated with the discreteness scale.

To demonstrate the emergence of a gravitational interaction about the mean-field vacuum we would reproduce the analysis of Refs.~\cite{Carone:2016tup,Batz:2020swk} with dimensional regularization replaced by the stochastic description. A detailed analysis will appear elsewhere. However, we briefly comment on the  scales associated with the emergent gravitational interaction. 
Refs.~\cite{Carone:2016tup,Batz:2020swk} identify the gravitational interaction by evaluation of the 2$\rightarrow$2 scattering amplitude at leading order in $1/N$. The relevant chains of loop integrals that arise are summed by evaluating a kernel proportional to $\langle {\cal T}_{\mu\nu}(x){\cal T}_{\alpha\beta}(y)\rangle$, where ${\cal T}_{\mu\nu}$ is the energy-momentum tensor for the free field in the vacuum spacetime.
With the stochastic regulator, the leading divergences of the integral are ${\cal O}(N/\tau^4)$ and ${\cal O}(Nq^2/\tau^2)$. These divergences lead to a Planck mass $M_{\rm Pl}\sim \sqrt{N}/\tau$. 

The effective Planck mass is hierarchically larger than the regulator scale, so  the effective quantum field theory description should  give way to smooth field evolution before dangerous Planck-scale effects are manifested. This also suggests the suppression  of higher-dimension operators in the effective theory at all scales.

\section{Conclusions}
We have presented a first analysis of an approach to quantum gravity in which discreteness of random fluctuations in a stochastic representation of quantum field theory provides a physical UV regulator. We found that, with a tuning of the constant in the potential, there is a self-consistent vacuum in which the expectation value of the composite metric operator fluctuates about a flat metric. We demonstrated that the conserved energy and momentum defined through an analogy to Noether's theorem for stochastic mechanics vanish identically, as expected in a metric-independent theory. We argued that fluctuations about the vacuum should give rise to an effective gravitational interaction as in prior analyses of composite gravity models that used standard nonphysical UV regulators as a proxy for a physical regulator like that provided by the stochastic field theory.

The absence of gravitation at ultra-short distances might be relevant for understanding the apparent low entropy of the early universe \cite{Penrose}. Furthermore, the possibility of a far-from-equilibrium initial field configuration leads to the possibility of imprints in early cosmology and on the horizon scale, for example by analogy with the prediction of a decreased power in the low-$\ell$ moments of the CMB power spectrum due to nonequilibrium effects in Bohmian field theory as demonstrated by Valentini
\cite{Valentini:2008dq}. Transitions between atomic states might also lead to temporary far-from-equilibrium configurations that 
would suppress further transitions when probed with ultrafast sequences of laser pulses tuned to atomic transition frequencies. 

As a nascent framework for quantum gravity, it will be important to revisit the puzzles of semiclassical black hole physics, spacetime singularities and cosmology. As gravitation arises only at distances longer than the discreteness scale in stochastic composite gravity, we do not necessarily expect or require that a microscopic counting of degrees of freedom would reproduce the Bekensein-Hawking black hole entropy. Cosmological spacetimes would need to be described by nonperturbative solutions to the stochastic field equations, but may be amenable to a self-consistent mean-field analysis as was presented here for a flat-space vacuum. It may prove interesting to consider how the spectral dimension of the theory varies with scale \cite{Carlip:2017eud}. It will also be necessary to extend the stochastic framework to include fermions and gauge fields. To this end, we might look towards the stochastic quantization framework of Parisi and Wu, which includes fermions and gauge fields \cite{Parisi:1980ys,Damgaard:1983tq}, for guidance.

\begin{acknowledgments}  
This work was supported in part by the NSF under Grant PHY-1819575 and by a Plumeri Award for Faculty Excellence at William \& Mary. \end{acknowledgments}

\end{document}